\documentclass[conference]{IEEEtran}
\IEEEoverridecommandlockouts

\usepackage{cite}
\usepackage{amsmath,amssymb,amsfonts}
\usepackage{algorithmic}
\usepackage{graphicx}
\usepackage{textcomp}
\usepackage{xcolor}
\def\BibTeX{{\rm B\kern-.05em{\sc i\kern-.025em b}\kern-.08em
    T\kern-.1667em\lower.7ex\hbox{E}\kern-.125emX}}

\usepackage{listings}
\usepackage{xcolor}
\lstdefinestyle{result}{
    frame=tb,     basicstyle=\ttfamily\scriptsize,     captionpos=b,     breaklines=true,
    breakatwhitespace=true,
    commentstyle=\color{gray},
    showstringspaces=false,
}

\newcommand{\ASIReq}{\textit{availableSpectrumInquiryRequest}}
\newcommand{\ASIResp}{\textit{availableSpectrumInquiryResponse}}

\begin{document}

\title{GPS Spoofing Attacks on Automated Frequency Coordination System in Wi-Fi 6E and Beyond
}

\author{
    \IEEEauthorblockN{Yilu Dong\IEEEauthorrefmark{1}, 
    Tianchang Yang\IEEEauthorrefmark{1}, 
    Arupjyoti Bhuyan\IEEEauthorrefmark{2},
    and Syed Rafiul Hussain\IEEEauthorrefmark{1}
}
\IEEEauthorblockA{\IEEEauthorrefmark{1}Department of Computer Science and Engineering, The Pennsylvania State University, University Park, PA, USA}
\IEEEauthorblockA{\IEEEauthorrefmark{2}INL Wireless Security Institute, Idaho National Laboratory, Idaho Falls, ID, USA}
yiludong@psu.edu, tzy5088@psu.edu, arupjyoti.bhuyan@inl.gov, hussain1@psu.edu
}

\maketitle

\begin{abstract}
The 6\,GHz spectrum, recently opened for unlicensed use under Wi-Fi\,6E and Wi-Fi\,7, overlaps with frequencies used by mission-critical incumbent systems such as public safety communications and utility infrastructure. To prevent interference, the FCC mandates the use of Automated Frequency Coordination (AFC) systems, which assign safe frequency and power levels based on Wi-Fi Access Point (AP)-reported locations. In this work, we demonstrate that GPS-based location reporting, which Wi-Fi APs use, can be spoofed using inexpensive, off-the-shelf radio equipment. This enables attackers to manipulate AP behavior, gain unauthorized spectrum access, cause harmful interference, or disable APs entirely by spoofing them into foreign locations. We validate these attacks in a controlled lab setting against a commercial AP and evaluate a commercial AFC system under spoofed scenarios. Our findings highlight critical gaps in the security assumptions of AFC and motivate the need for stronger location integrity protections. 

\end{abstract}


\section{Introduction}
With the increasing number of Wi-Fi devices, the existing 2.4\,GHz and 5\,GHz bands have become increasingly congested, leading to interference, degraded performance, and limited throughput. 
To address these limitations and meet the growing demands of modern wireless applications, the 6\,GHz spectrum (ranging from 5.925\,GHz to 7.125\,GHz) was opened for unlicensed use by
the Federal Communications Commission (FCC) in 2020~\cite{fcc:20-51}. Wi-Fi devices under 802.11ax (Wi-Fi\,6E) \cite{wifi6} and 802.11be (Wi-Fi\,7) \cite{wifi7} is allowed to operate in this spectrum. 
However, these newly available bands overlap with frequencies already used by incumbent licensed devices, many of which support mission-critical infrastructure. These systems include fixed microwave links for cellular backhaul, emergency services (e.g., police, fire, and medical communication networks), and utility telemetry for smart grids. Uncoordinated transmissions from unlicensed Wi-Fi Access Points (APs) operating in the same spectrum could cause harmful interference, potentially disrupting essential services and leading to severe consequences.

To enable safe coexistence between incumbent services and unlicensed Wi-Fi use in the 6\,GHz band, the FCC mandates the use of the Automated Frequency Coordination (AFC) system for outdoor, standard-power APs. The AFC is a cloud-based coordination mechanism, operated by certified providers, that determines which frequencies an AP can safely use and at what power levels. The AP must report its geographic location (e.g., GPS coordinates and height), and the AFC uses this information, along with a propagation model and a database of protected incumbents, to assess potential interference. The AFC then responds with an approved list of frequencies and transmission power levels tailored to the AP’s location. APs are required to comply with these guidelines before operating.

Given the reliance on location data, GPS-based positioning of AP is generally preferred over manually entered values to reduce the risk of human error or tampering. However, in this work, we demonstrate that even GPS-based location reporting can be spoofed using readily available radio equipment, leading to significant implications. 
An attacker with access to the AP or operating in its proximity can manipulate its reported location to obtain unauthorized access to protected frequencies or transmit at higher power levels, bypassing AFC safeguards. This may result in harmful interference to incumbent services and potential disruptions to critical infrastructure, posing national security risks.

Since APs periodically refresh their AFC permissions (e.g., every 24 hours), attackers can perform GPS spoofing near benign APs to alter their frequency and power allocations during these refreshes. 
At scale, this could lead to large-scale interference and may affect the operation of smart grid devices, leading to a potential outage \cite{he2016cyber}. 
GPS spoofing can also be used to disable benign APs by spoofing their location into foreign locations where no transmission is permitted. Given recent demonstrations of GPS spoofing via drones \cite{tibaldo2025gnss}, such attacks can closely mimic GPS signals from satellites and be launched remotely without requiring physical access to the AP.

In a controlled lab setting, we successfully spoofed GPS signals to manipulate a commercial AP’s reported location. We verified that this manipulation allows the AP to request arbitrary frequency and power configurations, or be rendered inoperative when placed virtually in a foreign area. Furthermore, using the ability to report arbitrary location parameters to AFC servers, we evaluated a commercial AFC system under various spoofed scenarios. While the system conformed to FCC-defined test cases, our findings reveal opportunities for further evaluation of edge cases and robustness under adversarial conditions.

In summary, our key contributions are:
\begin{itemize}
\item We perform the first security analysis of the AFC system, focusing on its reliance on GPS-based location reporting.
\item We design GPS spoofing attacks on AFC systems and analyze their potential impacts.
\item We validate the feasibility of GPS spoofing against a commercial Wi-Fi AP operating in the 6\,GHz band.
\item We evaluate AFC server implementations under spoofed inputs and discuss opportunities for robustness testing.
\end{itemize}

\noindent\textbf{Responsible Disclosure.} We have reported the discovered vulnerabilities to HPE and are working with them to improve the security of the products. Our experiments were done in a controlled environment that did not affect commercial AFC servers.
\section{Background}

\subsection{Fixed Service (FS) Link}
Fixed Service (FS) links refer to point-to-point or point-to-multipoint wireless communication systems that are deployed in fixed locations. These links form a critical part of national communication infrastructure, supporting high-capacity data transmission for applications such as Wireless Internet Service Provider (WISP) backhaul, mobile network backhaul, public safety communications, and telemetry for utilities and critical infrastructure. In the United States and many other countries, FS links are licensed to operate in the 6\,GHz spectrum. As these systems were already operating in the band before the introduction of unlicensed Wi-Fi use, they are referred to as \textit{incumbent} users of the spectrum. 

\subsection{Standard Power Access Points}
Standard Power Access Points (APs), introduced with Wi-Fi\,6E, are designed to operate in the newly opened 6\,GHz spectrum and support both indoor and outdoor deployments. These APs transmit at higher power levels compared to low-power indoor APs (Maximum Equivalent Isotropically Radiated Power (EIRP) at 36\,dBm), enabling extended coverage and the ability to serve more clients with a stronger and more reliable Wi-Fi signal. However, this increased transmission power also raises the risk of causing interference to other devices (e.g., incumbent systems) operating in the same band.
To address this risk, FCC requires all Standard Power APs to coordinate with an AFC system before operating on the 6\,GHz band. While prior work has evaluated the passive interference risk from Wi-Fi\,6E deployments~\cite{dogan2025evaluation}, it has not considered the possibility of active attacks where an adversary deliberately manipulates the AP’s behavior to cause harmful interference. In this work, we focus exclusively on Standard Power APs and use the term AP to refer to them for simplicity.

\subsection{Automated Frequency Coordination (AFC)}
\begin{figure}[h]
    \centering
    \includegraphics[width=\linewidth]{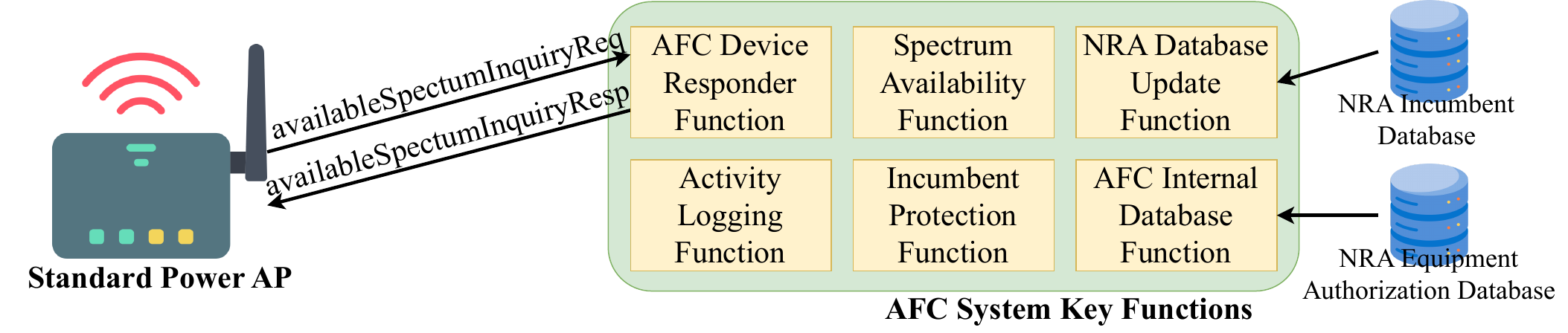}
    \caption{AFC System Architecture}
    \label{fig:AFC}
\end{figure}

The Automated Frequency Coordination (AFC) system is a cloud-based service designed to facilitate safe coexistence between unlicensed Wi-Fi devices and licensed incumbent systems in the 6\,GHz spectrum. Its primary role is to determine which channels an AP can use, and at what power levels, to avoid causing harmful interference to incumbent users such as FS links. A properly designed AFC system shall ensure the interference-to-noise ratio ($I/N$) is less than -6\,dB for all protected devices. 
Figure~\ref{fig:AFC} illustrates the overall architecture of the AFC system. To request authorization, a Standard Power AP sends an \ASIReq{} message to an AFC provider. This message includes the AP’s geographic location (typically obtained via GPS), device parameters, and other required metadata. Upon receiving the request, the AFC server queries databases maintained by the National Regulatory Authority (NRA) to retrieve information about protected incumbent systems in the area. It then applies standardized propagation models to calculate the maximum permissible transmission power across each 6\,GHz channel.
The server responds with an \ASIResp{} message that includes the approved channel and power combinations. The AP must comply with this configuration and may only transmit on channels explicitly authorized by the AFC.

\section{Security Analysis of The AFC System}

If AFC systems are compromised, the Standard Power 6\,GHz Wi-Fi APs will not get the correct channel and power allocation. This can lead to denial-of-service of 6\,GHz Wi-Fi clients or potential interference with other spectrum users if the received power exceeds the incumbent protection requirements. To identify the possible attack vectors, we present a security analysis of the AFC system. 

\subsection{General Security Requirements}
\label{subsec:sec_req}
From the AFC system requirements (WINNF-TS-1014) \cite{winnf-ts-1014} and the federal regulation 47 CFR \S~15.407(k) \cite{cfr:47:15.407}, we identify the following security requirements: 

\begin{itemize}
    \item \textbf{(REQ1)} The communication between the AP and the AFC server must be mutually authenticated, encrypted, and integrity-protected.
                \item \textbf{(REQ2)} Unauthorized users must not be able to access or modify internal AFC databases, including the list of protected incumbent systems.
    \item \textbf{(REQ3)} The AFC server must accurately compute and return the allowed frequency and power levels based on the AP’s parameters to ensure incumbent protection.
\end{itemize}

These guarantees are typically enforced via Transport Layer Security (TLS) for all AP-to-server communication~\cite{afc_spec}. Assuming the AP operates correctly and its TLS stack is uncompromised, attackers cannot intercept or tamper with AFC messages or spoof server responses.

\subsection{Limitations and Attack Surface}
The AFC server makes its core decisions (e.g., allowable frequencies and power) based almost entirely on the AP's location and device metadata. 
While these requirements secure the AFC communication interface and protect backend databases, they implicitly assume the AP-provided input (e.g., location) is trustworthy. This design choice introduces a subtle but critical vulnerability: if an attacker can manipulate the AP's reported parameters like location information, they can indirectly subvert the AFC system without breaking any cryptographic protections.

This input-based vulnerability is particularly concerning because many commercial standard power APs determine their geographic location using onboard GPS receivers. These GPS modules are often treated as trusted sensors and lack defenses against spoofing attacks. In practice, GPS signals can be spoofed using readily available, low-cost hardware. An attacker in proximity to an AP (or even remotely, via drone-based spoofing \cite{tibaldo2025gnss}) can inject falsified GPS signals to trick the AP into reporting a fake location to the AFC server.
This bypasses traditional security assumptions: although the AP and AFC server are communicating over a secure channel, the server is making decisions based on attacker-controlled input. 

\subsection{Threat Model}
In this paper, we consider that the AP is operating unaltered and has a secure connection (e.g., TLS) to communicate with the AFC server. Through a GPS spoofer, an attacker can send fabricated GPS signals and control the geographic location computed from the GPS receiver inside the AP. 
Since the received power from the GPS receiver is extremely low, an attacker can use a Software-Defined Radio (SDR) (e.g., the \$199 Flipper Zero \cite{flipperzero}) for the attack. The attack can be launched remotely with a long-range directional antenna or drones with a GPS spoofer. 

\subsection{Overview of GPS Spoofing}

The Global Positioning System (GPS) provides high-precision location and time synchronization. A receiver determines its position by calculating the travel time of unique, time-stamped signals from multiple satellites. By using data from at least four satellites, it can establish its three-dimensional location and the precise time. Today's civilian GPS is accurate to within 5 meters and 30 nanoseconds of Coordinated Universal Time (UTC) \cite{gpsgov2022accuracy}.

However, the civilian GPS signals are not authenticated. That opens a door for GPS Spoofing attacks. Anyone can generate GPS signals and transmit them to a GPS receiver. The receiver cannot differentiate between the signal coming from a valid satellite and the signal coming from an attacker. Also, since the signals are transmitted from the satellite, the received power on the ground is extremely low ($<$ -100 dBm) \cite{akos1996design}, making overshadowing the legitimate GPS signals easy. An attacker can use a cheap commercial-off-the-shelf SDR to transmit the GPS signals and perform an attack. 

\begin{figure}[h]
    \centering
            \includegraphics[width=\linewidth, trim=150 1050 150 0]{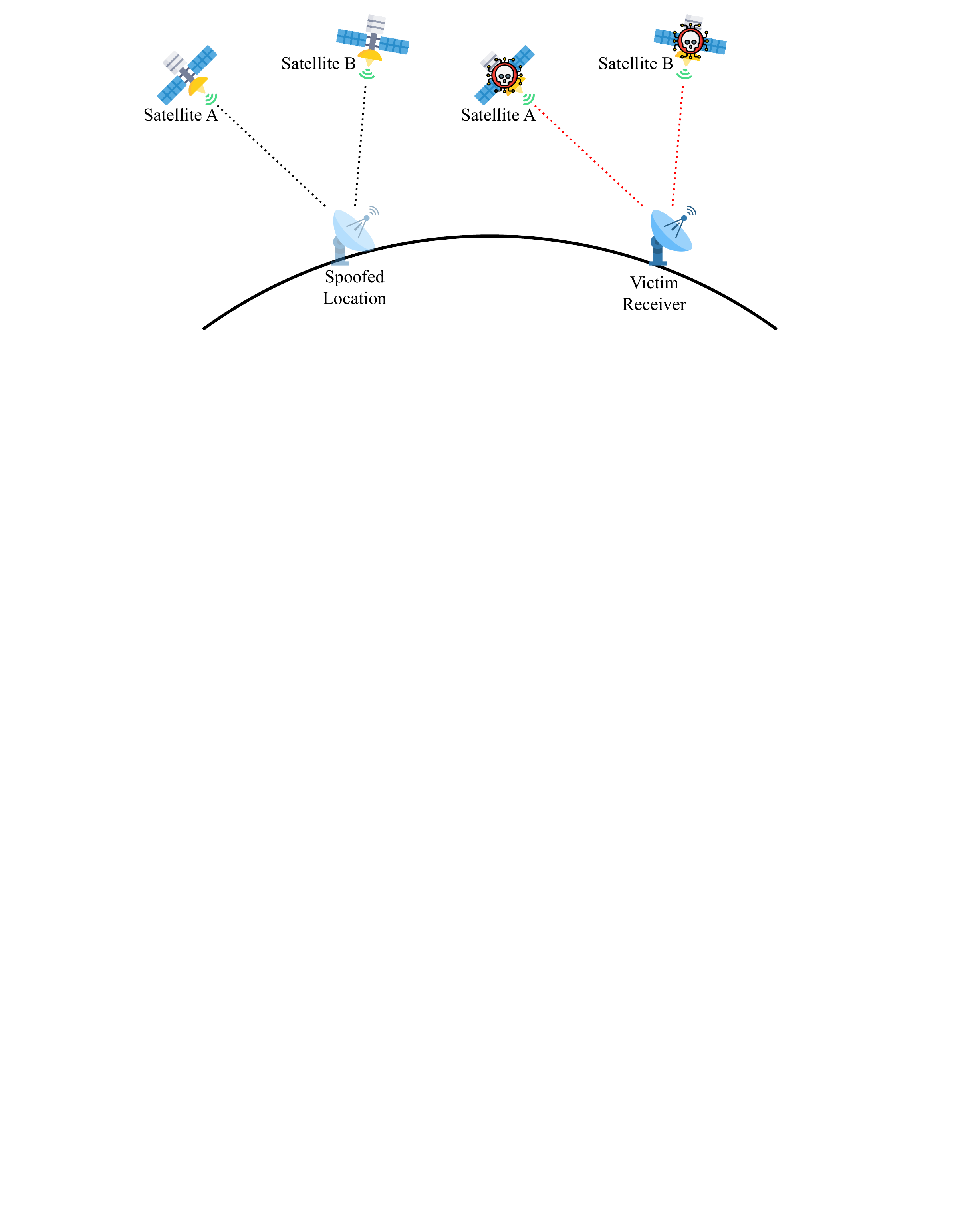}
    \caption{GPS Spoofing Attack}
    \label{fig:GPS_spoofing}
\end{figure}

As shown in Figure \ref{fig:GPS_spoofing}, an attacker can simulate the received GPS signals for an arbitrary location and use a specialized transmitter or a SDR to transmit the signal to the victim GPS receiver. With the received spoofing signals, the victim receiver will calculate its location as the spoofed location. For example, in the spoofed location, the receiver is expected to receive GPS signals from satellites A and B. The attacker generates these signals and sends them to the victim receiver. Since the victim receives the GPS signal from satellites A and B, it determines that it is at the spoofed location. 

Recent works propose advanced GPS spoofing attacks, making it more difficult to detect \cite{tippenhauer2011requirements}, covering a wider area \cite{tibaldo2025gnss}, and cheaper to implement \cite{cheng2025distributed}. GPS spoofing attacks will remain a vital threat to many commercial devices for a long time, including the newly introduced AFC system. 

\subsection{GPS Spoofing on AFC} \label{subsubsec:afc_attack}

\begin{figure}[h]
    \centering
    \includegraphics[width=\linewidth]{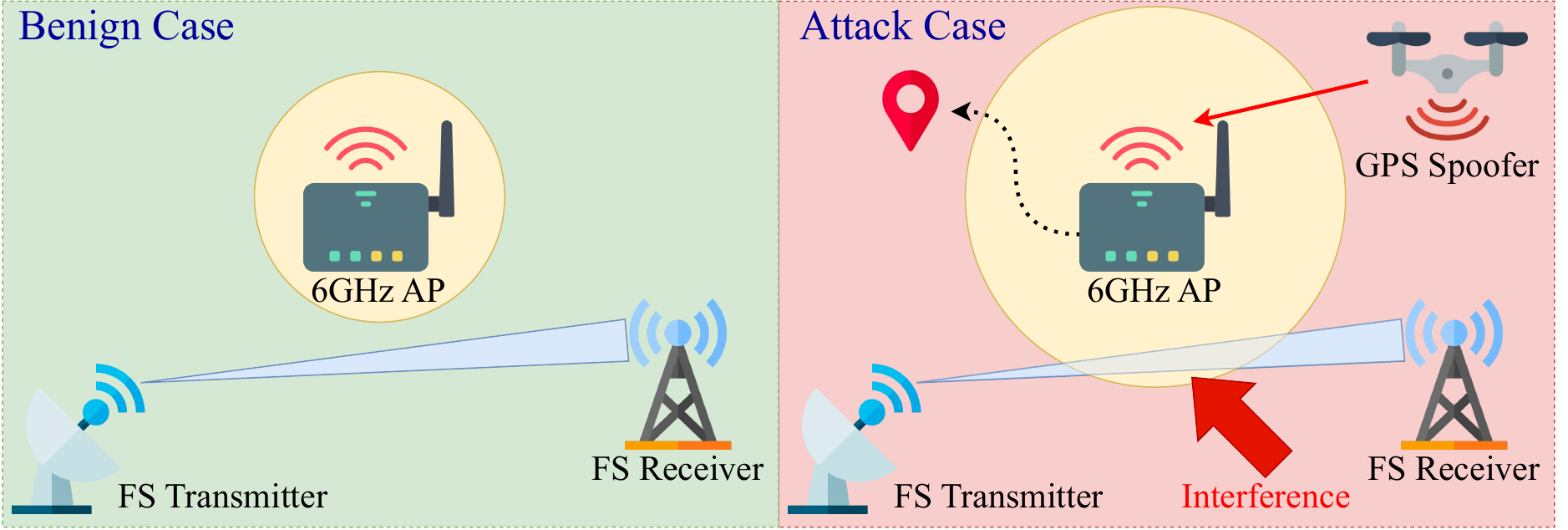}
    \caption{AFC Attack on Fixed Service (FS) Links}
    \label{fig:attack}
\end{figure}

Figure \ref{fig:attack} provides an example of a GPS spoofing attack on a 6\,GHz AP. Here in the benign case, the AP is connected to the AFC system and receives the correct available frequencies and associated power. As shown in the figure, the AP's transmission power is limited and does not cause interference to the existing link between the FS transmitter and receiver. However, in the attack scenario, the attacker can use a GPS spoofer to transmit fake GPS signals and let the AP calculate the wrong location. Then, the AP reports the spoofed coordinate to the AFC server and gets a higher allowed power. Now the AP can transmit the 6\,GHz Wi-Fi signal with a higher power, and the signal coverage increases. Therefore, the interference occurs and can interrupt the communication of the FS link. In addition, the attacker can also disable the 6\,GHz transmission on the AP by spoofing a foreign location or jamming the GPS signal. Without a valid coordinate, the AFC system cannot assign channels and the associated power to the AP. 

Some AFC implementations (including the system we tested) may enforce an extra check on the GPS timestamp. A naive spoofer that cannot send spoofing signals in real-time may not work in this case. To address this issue, we generate the estimated spoofing samples with a future timestamp and transmit the generated samples at the exact time of the timestamp used. After the setup, we observe that our AP receives the spoofing signals and sends the AFC request with the spoofed location. 

\subsection{Time-based Attacks}
Other than the GPS coordinates, the AP also requires a reliable time source for AFC operation. The AFC regulation \cite{cfr:47:15.407} mandates that the AFC client must update its information to the server at least once per day. Otherwise, it should stop the transmission. If the attacker can control the time of the system, it may roll back the system time to make the existing request never expire, or move forward the time to invalidate the current channel availability. Depending on how the AP obtains the time, the attacker may control the GPS time from spoofing or launch attacks on the Network Time Protocol (NTP).
\section{Experiment on Commercial AP}

\subsection{Experiment Setup}

\begin{figure}[t]
    \centering
    \includegraphics[width=\linewidth]{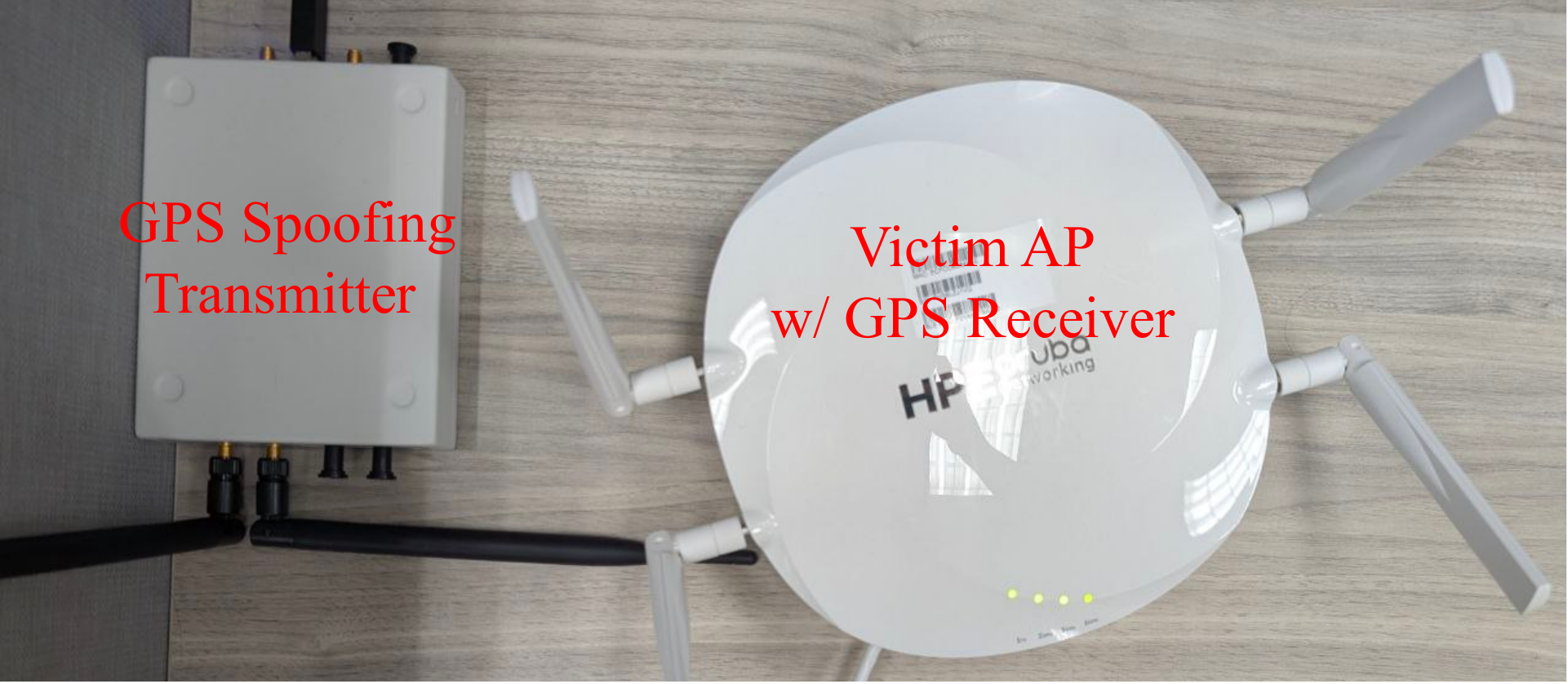}
    \caption{Experiment Setup}
    \label{fig:experiment}
\end{figure}

We conducted GPS spoofing attacks on an HPE Aruba AP-634 \cite{ap634}, which has Wi-Fi 6E and a built-in GPS receiver, using GPS-SDR-SIM \cite{gps-sdr-sim} with a USRP B210 \cite{b210}. All experiments were performed in a controlled environment, affecting only the AP, as shown in our setup in Figure \ref{fig:experiment}.

\subsection{Security Measures in Aruba AP} \label{subsec:aruba_sec}
During our experiments, we identified the following security measures in the Aruba AP we tested and found them compliant with the security requirements discussed in Section~\ref{subsec:sec_req}. 
\begin{itemize}
    \item {(M1)} All the communication between the AP and the AFC server (and management server) is protected by Transport Layer Security (aligned with REQ1). 
    \item {(M2)} The firmware of the AP is not publicly available, and the low-level control of the AP is not available to a normal user (aligned with REQ2). 
    \item {(M3)} For benign requests, a correct list of frequency and power allocation is returned. AFC requests with an incorrect GPS timestamp are not accepted (aligned with REQ3).
\end{itemize}

However, these security measures are not enough to prevent our GPS spoofing attack, and we can launch the attack without disabling any of the security measures mentioned. 

\subsection{\textbf{A1}: Interference Attack}

We find 3 attacks with our GPS spoofing tool. In these attacks, we first transmit the spoofing signal to the AP, wait until the AP collects enough location samples and calculates its location uncertainties, and then read the received channel availability from the AP. Here we provide a detailed explanation about the attack steps, observed results from the AP, and the impacts caused by these attacks. 

\subsubsection{Location Spoofing}

After the AP receives the spoofing signal, it calculates its location and location uncertainty in an ellipse form. Listing \ref{lst:ellipse} is an example output from the AP console. In this experiment, we generate spoofing signals for the coordinate (30.086965, -101.103761) (a rural area in Texas), and the AP calculated location is at coordinate (30.087050, -101.103714), 10 meters from the spoofed location. The error is likely introduced by an inaccurate clock in our transmitter. 

\begin{lstlisting}[style=result, caption={GPS ELLIPSE Information}, label={lst:ellipse}]
GPS ELLIPSE Information
-----------------------
Field       Value
-----       -----
latitude    30.087050
longitude   -101.103714
major-axis  18.052600
minor-axis  1.665372
angle       115.498839
time        2025-06-20 04:57:53
\end{lstlisting}

\subsubsection{Receiving AFC Channels Available}

After the AP calculates its location from GPS signals, it sends the \ASIReq{} to the AFC server and receives the response from \ASIResp{}. Listing \ref{lst:channels} provides a list of received AFC channels from the AP, including all the possible 6\,GHz Wi-Fi channels permitted in the United States. The allowed channels are valid for 24 hours. 

\begin{lstlisting}[style=result, caption={Received AFC Channels of \textbf{A1}}, label={lst:channels}]
Received afc channels 
----------------------
PHY Type               Allowed Channels
--------               ----------------
6GHz                   1 5 9 13 17 21 25 29 33 37 41 45 49 
                       53 57 61 65 69 73 77 81 85 89 93 117 
                       121 125 129 133 137 141 145 149 153 
                       157 161 165 169 173 177 181
6GHz 40MHz             1 9 17 25 33 41 49 57 65 73 81 89 
                       121 129 137 145 153 161 169 177
6GHz 80MHz             1 17 33 49 65 81 129 145 161
6GHz 160MHz            1 33 65 129
6GHz 80+80MHz          None
6GHz 320MHz_1          1
6GHz 320MHz_2          33
Present time           2025-06-20 05:13:13
Expiry time            2025-06-21 05:10:00
Country code           US
AFC channel expired    No
AFC channel required   Yes
\end{lstlisting}

\subsubsection{Power Associated with Channels}

Listing \ref{lst:power} provides the channel and the associated transmission power in Equivalent Isotropically Radiated Power (EIRP). In this spoofed location, all the channels are associated with the maximum power allowed in the specification, 36.0 dBm. The AP can select from these channels and start transmission with a power under the specified power limit. 

\begin{figure*}[h]
\begin{lstlisting}[style=result, caption={Max EIRP of AFC Channel}, label={lst:power}]
Max EIRP of AFC channel  
20MHz channel       1    5    9   13   17   21   25   29   33   37   41   45   49   53   57   61   65   69   73   77   81  
Max Eirp          36.0 36.0 36.0 36.0 36.0 36.0 36.0 36.0 36.0 36.0 36.0 36.0 36.0 36.0 36.0 36.0 36.0 36.0 36.0 36.0 36.0
20MHz channel      85   89   93  117  121  125  129  133  137  141  145  149  153  157  161  165  169  173  177  181
Max Eirp          36.0 36.0 36.0 36.0 36.0 36.0 36.0 36.0 36.0 36.0 36.0 36.0 36.0 36.0 36.0 36.0 36.0 36.0 36.0 36.0 
40MHz channel       1    9   17   25   33   41   49   57   65   73   81   89  121  129  137  145  153  161  169  177  
Max Eirp          36.0 36.0 36.0 36.0 36.0 36.0 36.0 36.0 36.0 36.0 36.0 36.0 36.0 36.0 36.0 36.0 36.0 36.0 36.0 36.0 
80MHz channel       1   17   33   49   65   81  129  145  161  
Max Eirp          36.0 36.0 36.0 36.0 36.0 36.0 36.0 36.0 36.0 
160MHz channel      1   33   65  129  
Max Eirp          36.0 36.0 36.0 36.0 
320MHz_1 channel    1  
Max Eirp          36.0 
320MHz_1 channel   33  
Max Eirp          36.0
\end{lstlisting}
\end{figure*}

\subsubsection{Impact of the Interference Attack}

In our experiments, we successfully validated the attacks proposed in \ref{subsubsec:afc_attack}. The AP under attack is allowed to operate on all channels with the maximum possible transmission power. If the AP is located near a mission-critical FS link, the AP may interrupt the existing service. If the AP is placed near a radio observatory, the transmitted signal can affect the observational results. 

\subsection{\textbf{A2}: No Channel Availability Attack Using a Foreign Location}

In addition to the attack above, we also test two denial-of-service attacks to disable all the transmissions in the 6\,GHz bands. In \textbf{A2}, we spoof the AP to a foreign location. We used the coordinate (30, 120) located in China. China does not permit the use of 6\,GHz frequencies yet, and the AFC system is not available there. As a result, we could not get any channel availabilities from the AFC server. 

\begin{lstlisting}[style=result, caption={Received AFC Channels of \textbf{A2} or \textbf{A3}}, label={lst:no_channels}]
Received afc channels 
----------------------
PHY Type                  Allowed Channels
--------                  ----------------
6GHz                      None
6GHz 40MHz                None
6GHz 80MHz                None
6GHz 160MHz               None
6GHz 80+80MHz             None
6GHz 320MHz_1             None
6GHz 320MHz_2             None
Present time              2025-06-20 11:36:04
Expiry time               None
Country code              None
AFC channel expired       Yes
AFC channel required      Yes
\end{lstlisting}

\subsection{\textbf{A3}: No Channel Availability Attack Using Invalid Time}

In \textbf{A2}, we find that the AP does not receive a response when the timestamp in the spoofed location is not aligned with the current time. By default, GPS-SDR-SIM \cite{gps-sdr-sim} generates GPS signals from the beginning of a day. We transmit the generated signal to the AP, and it cannot obtain the channel availability. 

Listing \ref{lst:no_channels} shows an example console output from the AP. Both approaches prevent the AP from getting the channel availability and transmitting on the 6\,GHz frequencies. 

\subsection{Use GPS Spoofing to Test the AFC Server} \label{subsec:test}

Beyond its use in malicious attacks, GPS spoofing can also be a valuable tool for verifying the correct implementation of an AFC system. As an independent party, we used this technique to run the official test cases from the AFC specification, with a particular focus on the Special Incumbent Protection (SIP) scenarios. The SIP tests ensure that the AFC system enforces exclusion zones around U.S. radio observatories, preventing Wi-Fi signals from interfering with observational results by prohibiting transmissions between 6650 and 6675.2\,MHz. We spoofed our AP's location into these zones and confirmed that all AFC responses were compliant with the expected results.
\section{Discussions}

\subsection{Defenses against GPS Spoofing Attacks}

Although detecting GPS spoofing attacks is difficult, adding countermeasures still increases the difficulty of a successful attack and improves the overall resilience of the AFC system. Some defense mechanisms can be implemented without additional hardware. We discuss 3 possible defense mechanisms. 

\subsubsection{Geofencing}

Most of the Wi-Fi APs are deployed in a fixed location. Only small variations should occur in its location. Under this assumption, the AFC provider can set up a geofence \cite{rodriguez2014geofencing} around the expected deployment location of the AP. When the AP reports a location outside the expected area, the system could raise a warning and shut down its service. 

\subsubsection{Detection Using AP Group}

A multi-receiver array can detect a single antenna GPS spoofer \cite{jansen2016multi, liu2021stars}, because a single antenna GPS spoofer cannot retain the relative position of the receivers. The Wi-Fi APs are often deployed in groups, covering a large area. Thus, the group of APs under the same operator can form such a multi-receiver array. With the APs reporting their location, the AFC system can detect possible spoofing attacks by computing the difference in the GPS location distances and the expected distances in deployment. 

\subsubsection{Network-Assisted Location Attestation}

A more resilient defense against location spoofing attacks involves integrating multiple location sources rather than depending on a single one. Relying exclusively on a single GNSS location service makes an Access Point (AP) highly vulnerable. By contrast, if the AP triangulates its position using diverse services, including various GNSS systems, WLAN, and network-based location services, the complexity for an attacker increases significantly, as they would need to spoof all signals concurrently. This method is used in Android's Fused Location API \cite{googlefused} and also validated by recent studies \cite{liu2025guardian} using multiple localization sources to identify GPS spoofing.

\subsection{Future Works}

As we discussed in Section \ref{subsec:test}, GPS spoofing enables us to test the AFC system independently. In the future, we can extend the tests to more coordinates with different corner cases. For example, the AFC system needs to select different propagation models in its calculation based on the distance between the AP and the incumbent receiver that needs to be protected. However, since the \ASIResp{} only provides the calculated channel availability and the associated power, we cannot directly infer the model they use from the responses. We can apply differential testing on this problem. An open-source AFC implementation, OpenAFC \cite{openafc}, implements all propagation models required by the specification. By comparing the output between OpenAFC and other private AFC implementations, we can understand which propagation model they are using and ensure the correctness and robustness of the AFC systems.
\section{Conclusion}
The 6\,GHz band enables high-performance Wi-Fi connectivity, but its safe use depends on the integrity of the AFC system, which assigns frequencies and power levels based on AP-reported GPS location. While AFC communications are secured, we show that the system remains vulnerable to GPS spoofing, allowing attackers to manipulate AP-reported location, bypass spectrum restrictions, and disrupt incumbent services. We validated these attacks on a commercial AP, demonstrating that current protections do not defend against input-level manipulation. These findings expose a critical vulnerability in AFC's trust model and highlight the broader risks of insecure sensor inputs. To ensure the reliability of spectrum sharing, future designs must incorporate mechanisms for robust and verifiable location reporting.
\section*{Acknowledgments}
This work is supported by a research grant from the Department of Energy (DOE) Office of the Cybersecurity, Energy Security, and Emergency Response (CESER), in collaboration with Idaho National Lab (INL).

\bibliographystyle{IEEEtran}
\bibliography{references}

\end{document}